\documentclass{article}
\usepackage[utf8]{inputenc}
\usepackage{amsmath}
\usepackage{amssymb}
\usepackage{amsthm}
\usepackage{graphicx}
\theoremstyle{definition}

\title{Theoretical possibilities of head transplant}
\author{Santanu Acharjee\\
Department of Mathematics, Gauhati University, Assam, India\\
e-mail: sacharjee326@gmail.com}
\date{}

\begin{document}

\maketitle

\section*{Abstract:}  Recently, Zielinski and  Sokal [ Zielinski, P., Sokal, P. (2016). Full spinal cord regeneration after total transection is not possible due to entropy change. Medical Hypotheses, 94, 63-65] proposed one hypothesis by considering three assumptions against the spinal cord regeneration after total transection. Thus, their claims are concluding that head transplant is not possible. But, using theoretical justifications, we show that head transplant is possible without  any information loss. We design a spinal cord logic circuit bridge (SCLCB), which is a reversible logic circuit and thus, we show that there is no information loss. \\

{\bf 2020 AMS Classifications:} 92C50, 94C11.\\

{\bf Keywords:} Spinal cord regeneration, entropy, laws of thermodynamics, head transplant, logic gates, SCLGB.

\section{Background}
In 2013, Canavero \cite{1} announced that full head (or body) transplant of a human is possible.  Since then, Canavero and Ren \cite{2}  have been facing  criticisms. Often,  ideas of Canavero \cite{1} or related procedural developments by Canavero and Ren \cite{3} are considered to be medically impossible tasks and unethical. One may find many articles  on medical impossibilities as well as ethical issues related to head transplantation in human \cite{4,5,6}, but only a few groups are supporting possibilities of head transplantation in human \cite{7,8}. In short, Canavero's  HEAVEN (The head anastomosis venture project) with spinal linkage (project GEMINI) \cite{1} is a matter of discussion from various sides, in spite of Ren and Canavero's hope \cite{7}. \\

Amidst all the debates regarding  possibilities and impossibilities related to head transplantation in human, an article \cite{9} attracted our attention. In 2016, Zielinski and Sokal \cite{9} considered a hypothesis and proved that full spinal cord regeneration after total transection is not possible. Their hypothesis concludes that head transplant with cent percent success  rate is not possible.  Their hypothesis is cited below.\\ 

``{\it The hypothesis is that full spinal cord restoration after transection is not possible due to irreversible loss of information about its structure, needed to restore the connections.}"\\

To prove their hypothesis, Zielinski and Sokal \cite{9} considered following three assumptions:\\
\begin{itemize}
    \item There are two million of axons
in the pyramidal tract in the cervical region and these axons are important for spinal cord regeneration and restoration of adequate quality of life of a patient.
    
    \item  The second assumption is that the 
    regeneration of damaged spinal cord should lead to axonal growth through the lesion site from the proximal end of the cord to the distal end and their re-connections with adequate target cells with loss of distal parts of axons, below the level of transection and there is an equal number of targets.

   \item The axonal growth of the severed spinal cord is made fully possible
\end{itemize}

 Zielinski and Sokal \cite{9} provided some basic mathematical  justifications (using permutations) behind their claims, which cannot be neglected at first glance. They also claimed that there is a  lack of mathematical background in this area of research and thus unnecessary expenditures of high research funds. But, in next section, we  provide  mathematical justifications and prove that head transplant with cent percent success rate is possible in theoretical sense. \\
\section{Hypothesis}

We consider the following hypothesis:\\

{\it Head transplant is possible without any loss of information. Moreover, a logic circuit (SCLCB) can be used to transmit nerve signals in case of transection of spinal cord due to spinal cord injury.}

\section{ Shannon entropy, uncertainty and information loss during head transplant}

In \cite{9}, Zielinski and Sokal considered that there are two millions pyramidal axons in the spinal cord. They considered spinal cord  as an open system whose entropy would be lost after injury. After transection due to injury, the proximal part and the distal part would be formed. Both of these parts would have new entropy. Now, while reconnecting  proximal part with distal part, if one makes incorrect re-connections, the brain  may have some problems to reorganise, refining its own connectivity due to the brain plasticity \cite{10}. Here, we develop related mathematics to show that head transplant is not possible in case of maximum uncertainty. Let us rename two millions pyramidal axons as $A_1, A_2, ..., A_{2,000,000}$. If $p_i$ be the probability of establishing  correct interconnection of a pyramidal axons $A_i$  from proximal part to its counter distal part, then $p_i=\frac{1}{2,000,000}$; $\forall i=1,2,..., 2,000,000$. In this case, if $H_{random}$ be the Shannon entropy \cite{11}, then $H_{random}=-\sum_{i=1}^{2,000,000}p_ilog_2(p_i)=log_2(2000000)$. Thus, maximum uncertainty is obtained when one wants to establish interconnections randomly.  We consider an ideal hypothetical condition where interconnection  after total  transection will be done as same as before total transection. In this case, if $p_i'$  be the probability of establishing  correct interconnection of a pyramidal axons $A_i$  from proximal part to its counter distal part, then $p_i'=1$; $\forall i=1,2,..., 2,000,000$. Let $H_{ideal}$ be the Shannon entropy \cite{11} for this ideal case, then $H_{ideal}=-\sum_{i=1}^{2,000,000}p_i'log_2(p_i')=0$. Thus,  the entropy is zero i.e. there is no uncertainty only in case of ideal hypothetical case. Zielinski and Sokal \cite{9} assumed random re-connections and thus obtained 2,000,000! permutations. But, why do we assume that neurosurgeons will establish random interconnections of pyramidal axons? In \cite{13}, Ren et al.  discussed various medical methods along with experimental evidences regarding possibilities of spinal cord fusion, axons regeneration, etc. Moreover, they quoted the following paragraph from Busy et al. \cite{14}.

``{\it The pyramidal tract is not essential to useful control of the skeletal
musculature. In the absence of the corticospinal fibers, other fiber systems, particularly some multi‑neuronal mechanism passing through the mesencephalic
tegmentum, are capable of producing useful, well‑coordinated, strong, and delicate movements of the extremities.}"

Thus, the above paragraph encourages us to develop mathematical possibilities with functionally equivalent mechanism. Let us subdivide 2,000,000 pyramidal axons into some classes based on their functional equivalence and non-functional equivalence mechanisms.  Let $F_1$, $F_2$,..., $F_K$ be $k$ classes having functional equivalence mechanism and $F_{K+1}$ is a class of pyramidal axons based on non-functional equivalence mechanism. We assume $|F_1|+|F_2|+...+|F_K|=2,000,000-n$ and $|F_{K+1}|=n$, where $n<<2,000,000$. Due to functional equivalence mechanism,  Shannon entropy  for class $F_i$ (symbolically $H_{F_i}$) is zero, for $i= 1, 2, 3, ..., k$ and $H_{F_{k+1}}$ = $log_2(n)$ due to non-functional equivalence. If $H_{functional}$ be the entropy based on functional equivalence and non-functional equivalence mechanisms, then due to Shannon \cite{9}, we obtain $H_{functional}$=$\sum\limits_{i=1}^{k+1}H_{F_{i}}$=$ log_2(n)$. Since $n<<2,000,000$, thus $H_{ideal}\leq H_{functional}<H_{random}$. Thus, it can be concluded mathematically that functions will be restored due to re-connections as long as the mismatch is not extreme \cite{9}. Since, entropy of information theory is connected to the  second law of thermodynamics \cite{15}, thus  we urged against the claim of  Zielinski and Sokal \cite{9}. Hence, we conclude that  spinal cord regeneration is theoretically possible  if neurosurgeons identify  classes $F_i$s with  functional equivalence mechanisms at most matching. \\

\section{Spinal Cord Logic Circuit Bridge (SCLCB)}

\begin{figure}
    \centering
    \includegraphics[width=5in, height=3in]{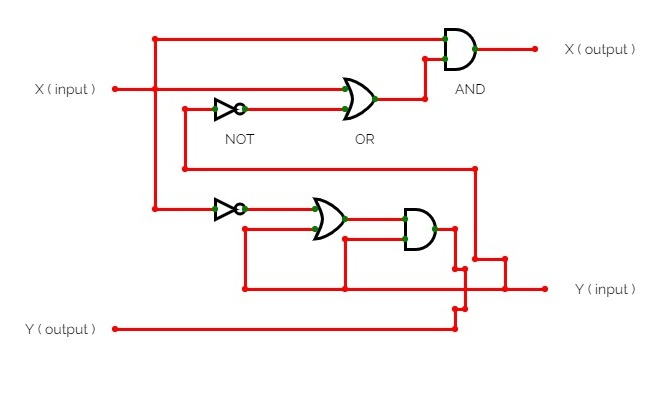}
    \caption{ Spinal Cord Logic Circuit Bridge (SCLCB)}
    \label{fig:1l}
\end{figure}

\begin{figure}
    \centering
    \includegraphics[width=5in, height=3in]{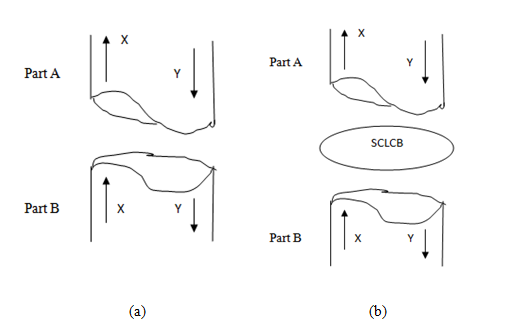}
    \caption{(a) Spinal cord injured (SCI) and transected (b)     SCLCB between transected spinal cord}
    \label{fig:2l}
\end{figure}

Recently, Nie et al.\cite{16} showed that nerve signal transmission after spinal cord injury is possible in quantum realm and thus, they proposed spinal cord
quantum bridge (SCQB). This inspires us to built a logic circuit using only three logical operators AND, OR, and NOT. Figure \ref{fig:1l} shows a logic circuit  which takes input X  and input Y in right and  left side respectively. Again, it gives output Y and output X  in right and  left side respectively. We call this logic circuit as Spinal Cord Logic Circuit Bridge (SCLCB).\\

In  (a) of figure \ref{fig:2l}, an injured spinal cord is shown, which is transected.   Here, X represents  transmission of  nerve signals in upward direction and 
Y represents transmission of  nerve signals in downward direction. More elaborately, X represents  
transmission of  sensory nerve signals from peripheral nerves to central nerves and  Y represents  transmission of motor nerve signals from central nerves to outer
peripheral nerves. But, due to transection of spinal cord, transmission of X and Y are  stopped.  In  (b) of figure \ref{fig:2l}, we implant SCLCB between  two breakpoints  Part A and Part B. While implanting  SCLCB, 
we are to connect X(input) and Y(output)  in Part B and  X(output) and Y(input)  in Part A of transected spinal cord. SCLCB has the property that it takes both the signals X and Y simultaneously as  inputs at time $t_1$ and transmit signals X and Y as outputs at time $t_2$. For example,  if we consider X=1 and Y=1 as inputs at time $t_1$ i.e. signals transmit in both directions   at time $t_1$,  then  we can  easily check that  SCLCB gives outputs X=1 and Y=1 at time $t_2$. One may check for other three cases of inputs viz. (i) X=1 and Y=0 (ii) X=0 and Y=1, and (iii) X=0 and Y=0. Thus, we propose that SCLCB can be implant between two parts of spinal cord after transection due to spinal cord injury. Thus, we propose that transmissions of nerve signals are  possible in SCLCB. Since, both SCLCB and SCQB \cite{16} show the possibilities to transmit nerve signals during spinal cord injury, thus we theoretically predict that head transplant is possible. \\

\section{Information loss in digital circuit and SCLCB}

Recently, H\"{a}nninen  et al. \cite{17} established methodologies to quantify irreversible information loss in digital circuits. The property of mapping a set of input bits onto  a set of outputs in a logical operation is called logical reversibility \cite{17}. In a logically reversible computation, the inputs can be known from the outputs. Physical reversibility occurs in an isolated system. According to  H\"{a}nninen  et al. \cite{17}, a logically irreversible computation can be obtained in an isolated system.  One may refer to Keyes and  Landauer \cite{20} for energy dissipation in logic gates. We cite the following paragraph from  H\"{a}nninen  et al. \cite{17}.\\

``{\it One bit can be in two states, so the associated entropy is $S={k_\beta}ln2$ Erasing an unknown bit, changing either 0 or 1 to a NULL state, means there is a transfer of this entropy to the environment with associated free energy:\\
\centerline{$\Delta E=TS={k_\beta}Tln2$.}

Thus a physical implementation of a bit ensure or any logical operation that loses 1 bit of information must necessarily dissipate an amount of heat greater than or equal to $\Delta E$.}"\\

In case of logical reversible computation,  if the computation can be made sufficiently slow, then dissipation of arbitrary small amount of energy is occurred \cite{21}. This claim is quite similar to the claim made by Ren et al. \cite{13} in favour of extremely sharp cut. Moreover, Bennett \cite{21} suggested that minimisation of  the energy  dissipation can be obtained near thermodynamic equilibrium. Thus, our ideas of functionally equivalent mechanism and $H_{functional}$ are important for neurosurgeons for head transplantation. Bennett \cite{21} showed that a logically irreversible computation can be made logically reversible in each step. But, it is important to note that SCLCB is reversible logical circuit because one can determine both the inputs from the outputs \cite{17}. Thus, there is no case of information loss. Hence, we can conclude that there is no dissipation of energy  in SCLCB \cite{20,21}. More suitable chemical and electronic technologies viz. molecular logic gates \cite{23}, electronic circuit  design \cite{22}, etc.  may be used  to make  SCLCB more effective. Recently, clinical trails on partial restoration of spinal cord neural continuity were done \cite{18, 19}. Thus, we should remain optimistic about  head transplant. So, we can justify our hypothesis.

\section{Conclusion}

Zielinski and  Sokal \cite{9} predicted that full spinal cord regeneration after total transection is not possible due to entropy change, and thus it concluded indirectly that  head transplant is impossible. To predict, they considered three assumptions. In this paper, considering the same assumptions we proved that head transplant is possible. We showed that functional equivalence mechanisms yield less information loss. Moreover, we design the spinal cord logic circuit bridge (SCLCB) as a technical measure for transmission of nerve signals through spinal cords after transection due to spinal cord injury. Both SCLCB and SCQB of \cite{16} show the possibilities of head transplantation in theoretical sense. Since, the procedure of head transplantation requires sophistication in medical and engineering technologies, thus we assume that the domain of head transplant attract attention of several interdisciplinary fields.\\

{\bf Competing interests:} The author has no conflict of interest.\\

{\bf Funding statement:}  No funding received.\\

{\bf  Ethical approval/Ethical approval:} Not required.\\

\end{document}